\documentclass[aps,superscriptaddress,showpacs,floatfix,preprint]{revtex4}
\usepackage{times}
\usepackage{xspace}
\usepackage{graphicx,morefloats}
\usepackage{color}
\usepackage{amsmath, amsthm, amssymb}

\begin{document}
\title{Regime shifts in models of dryland vegetation}
\author{Yuval R. Zelnik}
\author{Shai Kinast}
\author{Hezi Yizhaq}
\author{Golan Bel}
\affiliation{Department of Solar Energy and Environmental Physics, Blaustein Institutes for Desert Research, Ben-Gurion University of the Negev, Sede Boqer Campus 84990, Israel}
\author{Ehud Meron}
\affiliation{Department of Solar Energy and Environmental Physics, Blaustein Institutes for Desert Research, Ben-Gurion University of the Negev, Sede Boqer Campus 84990, Israel}
\affiliation{Department of Physics, Ben-Gurion University, Beer
Sheva, 84105, Israel}

\begin{abstract}
Drylands are pattern-forming systems showing self-organized vegetation patchiness, multiplicity of stable states and fronts separating domains of alternative stable states. Pattern dynamics, induced by droughts or disturbances, can result in desertification shifts from patterned vegetation to bare soil. Pattern-formation theory suggests various scenarios for such dynamics; an abrupt global shift involving a fast collapse to bare soil, a gradual
global shift involving the expansion and coalescence of bare-soil domains, and an incipient shift to a hybrid state consisting of stationary bare-soil domains in an otherwise periodic pattern. Using models of dryland vegetation we address the question which of these scenarios can be realized. We found that the models can be split into two groups: models that exhibit multiplicity of periodic-pattern and bare-soil states, and models that exhibit, in addition, multiplicity of hybrid states. Furthermore, in all models we could not identify parameter regimes in which bare-soil domains expand into vegetated domains. The significance of these findings is that while models belonging to the first group can only exhibit abrupt shifts, model belonging to the second group can also exhibit gradual and incipient shifts. A discussion of open problems concludes the paper.


\end{abstract}

\maketitle

\section{Introduction}
Water-limited landscapes can generally be described as mosaics of vegetation and bare-soil patches of various forms. Increasing empirical evidence supports the view that this type of vegetation patchiness is a self-organization phenomenon that would have occurred even in perfectly homogeneous physical environments~\cite{Valentin1999catena,Deblauwe2008geb}. Much insight into the mechanisms that drive self-organized vegetation patchiness has been achieved using mathematical models of water-limited landscapes~\cite{Gilad2007jtb,Rietkerk2008tree,Borgogno2009rev_geo}. These models first demonstrate that uniform vegetation states can go through spatial instabilities to periodic vegetation patterns upon increasing environmental-stress parameters. They further highlight two main feedbacks that are capable of producing such instabilities~\cite{Meron2012eco_mod}. The first is a positive feedback between biomass and water that develops as a result of an infiltration contrast between bare and vegetated areas (infiltration feedback). The second is a positive feedback between above-ground and below-ground biomass, related to the root-to-shoot ratio, a characteristic trait of any plant species (root-augmentation feedback).
Model studies of vegetation pattern formation along a rainfall gradient have revealed five basic vegetation states~\cite{vonHardenberg2001prl,Rietkerk2002an,Lejeune2004ijqc,Gilad2004prl}: uniform vegetation, gap patterns, stripe (labyrinth) patterns, spot patterns and uniform bare-soil. Another significant result is the existence of precipitation ranges where alternative stable vegetation states coexist. These are generally bistability ranges of any consecutive pair of basic states: bare-soil and spots, spots and stripes, stripes and gaps and gaps and uniform vegetation. Within any bistability range, spatial mixtures of the two alternative stable states can form long transient patterns that culminate in one of the two alternative states, or stable asymptotic hybrid patterns~\cite{Meron2012eco_mod}.

\begin{figure}
\includegraphics[width=110mm]{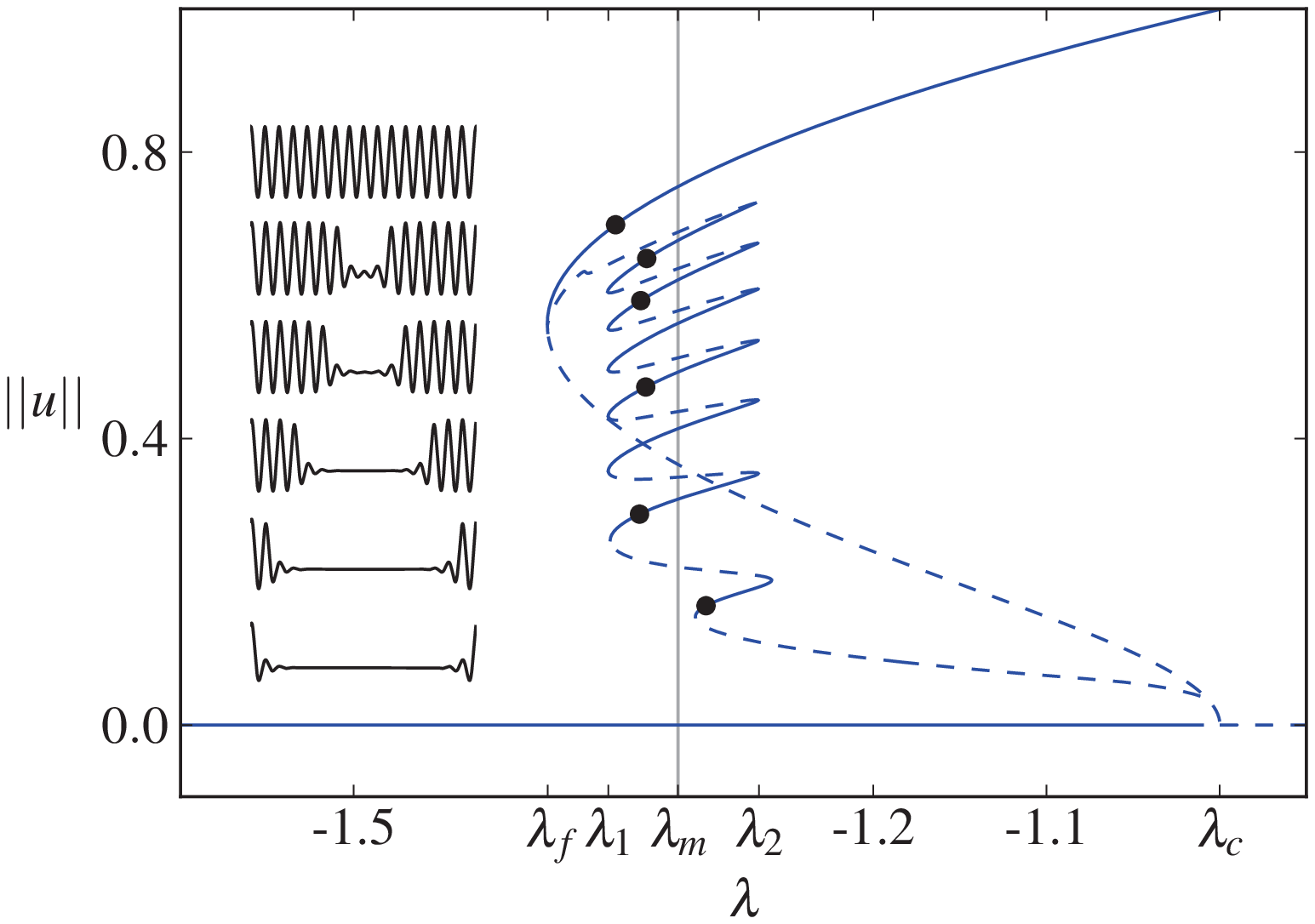}
\caption{
A bifurcation diagram showing bistability of a uniform zero state and a periodic pattern state, and some of the many hybrid states that may exist in such a case. Solid (dashed) lines denote stable (unstable) states. The hybrid states are described by solution branches that snake down towards the zero state and correspond to holes of increasing size in periodic patterns, as the insets on the left show. The horizontal axis represents a control parameter while the vertical axis represents a global measure of the state variable, such as the L2 norm. The vertical line denotes the Maxwell point $\lambda=\lambda_m$. The interval $\lambda_1<\lambda<\lambda_2$ is the snaking or hybrid-state range.  The diagram was calculated using the Swift-Hohenberg equation, a minimal model for bistability of uniform and patterned states~\cite{Knobloch2008nonlinearity}. From ~\cite{Bel2012theo_ecol}.
}
\label{fig:snaking}
\end{figure}

The mathematical theory of hybrid patterns is far from being complete. Much progress, however, has been made for the simpler case of bistability of uniform and periodic-pattern states, using simple pattern formation models such as the Swift-Hohenberg equation~\cite{Knobloch2008nonlinearity}. According to this theory a bistability range of uniform and patterned states may contain a subrange (or an overlapping range) of stable localized patterns, coexisting with the two alternative stable states. For bistability of bare-soil and vegetation spot patterns these localized patterns would correspond to isolated spot-pattern domains  in an otherwise bare-soil, or conversely, to isolated bare-soil domains in an otherwise periodic spot pattern. The appearance of these mixed-pattern or hybrid states can be understood intuitively by focusing on the dynamics of the transition zones that separate the two alternative stable states. These zones,
are fronts that can be stationary or propagating. In the case of bistability of two uniform states, isolated fronts always propagate, except for a singular control-parameter value, the so called Maxwell point, at which the direction of propagation changes~\cite{Bel2012theo_ecol}\footnote{A pair of fronts
propagating towards one another, however, can slow down and become stationary due to repulsive front interactions~\cite{Hagberg1994nonlinearity}.}. Bistability of uniform and pattern states, on the other hand, allow for an additional behavior; isolated fronts can be stationary or pinned in a \emph{range} of the control parameter~\cite{Pomeau1986pd}. Such a range can give rise to many hybrid states, because the fronts that constitute the boundaries of the alternative-state domains are stationary. In a diagram that shows the various states as functions of the control parameter, the hybrid states often appear as solution branches that "snake" down from the periodic-pattern branch towards the uniform (zero) state as Fig.~\ref{fig:snaking} illustrates. The control-parameter range where these solutions exists is often called the snaking range and the appearance of such solutions is described as homoclinic snaking~\cite{Knobloch2008nonlinearity}. In the following we will refer to this range as the "hybrid-state range" to allow for multistability of hybrid states that is not associated with homoclinic snaking.

Bistability of alternative stable states has been studied extensively in the context of ecosystem regime shifts, i.e. sudden transitions to a contrasting state in response to gradual changes in environmental conditions~\cite{Scheffer2001nature,Scheffer2003tree}. Such shifts have been observed in lakes, coral reefs, oceans, forests and arid lands. Global shifts from one stable state to another, however, may not necessarily be abrupt. Ecosystems are continuously subjected to local disturbances whose effects are spatially limited. Examples of such disturbances in the context of water-limited vegetation include clear cutting, grazing, infestation and limited fires.
These disturbances can induce fast \emph{local} transitions to the alternative stable state, but, according to pattern formation theory, the subsequent dynamics may proceed slowly by the expansion and coalescence of the domains of the alternative stable state through front propagation and front collisions. Such a succession of processes eventually leads to a global regime shift, but in a gradual manner~\cite{Bel2012theo_ecol}.

How slow can gradual shifts be? When the two alternative stable states are spatially uniform the pace of a gradual shift depends on the value of the control parameter relative to the Maxwell point; the larger the distance from that point the faster the gradual shift. This result often holds for bistability of uniform and patterned states too, except for one important difference - the value of the control parameter should be outside the hybrid-state range (but still within the bistability range)~\cite{Pomeau1986pd,Knobloch2008nonlinearity}. The difference between abrupt and gradual shifts can be dramatic, as Fig. \ref{fig:abrupt_vs_gradual} illustrates. For systems whose spatial extent is much larger than the size of a spot, gradual shifts can occur on time scales that are orders of magnitude longer than those of abrupt shifts.

Within the hybrid-state range global regime shifts are not expected to occur in steady environments. The system rather shows spatial plasticity; any spatial disturbance pattern shifts the system to the closest hybrid pattern, which is a stable stationary state and therefore involves no further dynamics. It is worth noting that transitions from the periodic pattern state to hybrid patterns, within the hybrid-state range, can also occur as a result of global uniform environmental changes, such as a precipitation drop or a uniform disturbance, provided the initial pattern is not perfectly periodic, e.g. hexagonal spot pattern containing penta-hepta defects~\cite{Meron2012eco_mod}.

Bistability of uniform and patterned states is most relevant to desertification, a regime shift involving a transition from a productive vegetation-pattern state to an unproductive uniform bare-soil state~\cite{vonHardenberg2001prl,Rietkerk2004science}. To what extent are the general results of pattern formation theory
displayed in Figs. \ref{fig:snaking} and \ref{fig:abrupt_vs_gradual} applicable to the specific context of desertification? We address this question by studying specific models of vegetation pattern formation of various degrees of complexity. The manuscript is organized as follows. In section \ref{sec:vegmodels} we briefly review the models of water-limited vegetation considered here.
In section \ref{sec:results} we present numerical results for these models, distinguishing between models for which we found indications for hybrid states (homoclinic snaking) and models for which we have not found such indications. These results are discussed and summarized in section \ref{sec:summary}.

\section{Models for spatial vegetation dynamics}\label{sec:vegmodels}
We chose to study several representative models of increasing complexity.
All models are deterministic and specifically constructed to describe vegetation patchiness in water-limited flat terrains (unlike the variant of the Swift-Hohenberg equation used to produce Figs. \ref{fig:snaking} and \ref{fig:abrupt_vs_gradual}). The degree of complexity is reflected by the number of dynamical variables and by the number of pattern-forming feedbacks the model captures. The models consist of partial differential equations (PDEs) for a continuous biomass variable and possibly for additional water variables, depending on the model. All models capture an instability of a uniform vegetation state to a periodic-pattern state and a bistability range of periodic patterns and bare soil.
%
\subsection{Lefever-Lejeune (LL) model}\label{sec:Lejeune}
The simplest model we consider is a single-variable model for a vegetation biomass density, $b\left({\bf{r}},t\right)$, introduced by Lefever and Lejune ~\cite{Lefever1997bmb,Lejeune1999}.
We chose to study a simplified version of this model~\cite{Lejeune2002pre,Lefever2000,Tlidi2008lnp} whose form in terms of non-dimensional variables and parameters is
\begin{align}
 \partial_t b=\left(1-\mu\right)b+\left(\Lambda-1\right)b^2-b^3+\frac{1}{2}\left(L^2-b\right)\nabla^2 b-\frac{1}{8}b\nabla^4 b\,. \label{TlidiEq}
\end{align}
In this equation the parameter $\mu$ is the mortality to growth ratio, $\Lambda$ is the degree of facilitative, relative to competitive, local interactions experienced by the plants, and $L$ is the ratio between the spatial range of facilitative interactions and the range of competitive interactions. The spatial derivative terms represent short range facilitation and long range competition, a well known pattern formation mechanism~\cite{Gierer1972kibernetic}.
The agents responsible for this mechanism in actual dryland landscapes are nonlocal feedbacks involving water transport towards growing vegetation patches. Explicit modeling of these feedbacks requires the addition of water variables.
Although the model does not include a precipitation parameter, water stress can be accounted for by increasing the mortality parameter $\mu$.
In what follows we will refer to this model as the LL model.
\subsection{Modified Klausmeier (K) model}\label{sec:Klausmeier}
Next in degree of complexity is a modified version~\cite{vanderStelt2012nonl_sci} of a model introduced by Klausmeier \cite{Klausmeier1999science}, hereafter the K model. In addition to a biomass density variable, $b$, this model contains a water variable, $w$, which we regard as representing soil-water content. The model equations, expressed in terms of non-dimensional quantities, are
\begin{subequations}
\begin{align}
  \partial_t b&=G(w,b)b-\mu b+\nabla^2 b, \label{Kb} \\
  \partial_t w&=p-w-G(w,b)b +D_w \nabla^2 w\,, \label{Kw}
\end{align}
\end{subequations}
where $G=wb$. According to equation \eqref{Kb}
the biomass growth rate, $G$, increases with the biomass density,
reflecting a positive local facilitation feedback. Natural mortality at a rate $\mu$, acts to reduce the biomass, and local seed dispersal or clonal growth,
represented by the diffusion term $\nabla^2b$, act to distribute the biomass to adjacent areas. The water dynamics (equation \eqref{Kw}) is affected by precipitation with a rate $p$, evaporation and drainage ($-w$), biomass-dependent water-uptake rate ($-Gb=-b^2w$),
and by soil-water diffusion. The pattern-forming feedback in this model is induced by the combined effect of higher local water-uptake rate in denser vegetation patches and fast water diffusion towards these patches, which inhibits the growth in the patch surroundings. This mechanism may be applicable to sandy soils for which $D_w$ is relatively large. This third type of pattern-forming feedback (besides the infiltration and root-augmentation feedbacks) has not been stressed in earlier studies.

The original Klausmeier model~\cite{Klausmeier1999science} does not include a
water diffusion term, but rather an advection term to describe runoff on a slope.
While accounting for banded vegetation on a slope, the original model does not produce stationary vegetation patterns in flat terrains. To capture the latter we added the soil-water diffusion term~\cite{vanderStelt2012nonl_sci}. Since we focus on plane terrains we do not need an advection term and therefore omitted it.

\subsection{Rietkerk et al. (R) model}\label{sec:Rietkerk}
The third model we consider, the R model, distinguishes between below-ground and above-ground water dynamics by introducing two water variables; $w$, representing soil water, and $h$, representing surface water. This three-variable model has been introduced by Rietkerk et al. \cite{Rietkerk2002an, HilleRisLambers2001ecology} and consists of the following non-dimensional equations~\footnote{The non-dimensional form of the equations was derived from the dimensional model, presented in \cite{Rietkerk2002an}, using the following transformation (the $\tilde{}$ denotes the variables in \cite{Rietkerk2002an}):
$t=\tilde{t}/t_0$, $x=\tilde{x}/x_0$, $b=P/k_2$, $w=W/k_1$, $h=O/k_1$, $D_w=\tilde{D}_wt_0/x_0^2=\tilde{D}_w/\tilde{D}_P$, $D_h=\tilde{D}_Ot_0/x_0^2=\tilde{D}_O/\tilde{D}_P$, $\mu=dt_0$, $\alpha=\tilde{\alpha}t_0$, $f=W_0$, $\gamma=k_2/(k_1c)$, $\nu=r_Wt_0$, $p=Rt_0/k_1$  where $t_0=1/(c g_max)$, $x_0=\sqrt{\tilde{D}_pt_0}$.}:
\begin{subequations}
\begin{align}
 \partial_t b&=G(w) b-\mu b+\nabla^2b\,, \label{Rb} \\
 \partial_t w&= Ih - \nu w -\gamma G(w) b +D_w\nabla^2 w\,, \label{Rw} \\
 \partial_t h&=p- Ih +D_h\nabla^2h\,, \label{Rh}
\end{align}
\end{subequations}
where
\begin{equation}
    G=\frac{w}{w+1}\,,\qquad I=\alpha\frac{b+f}{b+1}\,.\label{GI}
\end{equation}
In equation \eqref{Rb} the biomass growth rate, $G=G(w)$, depends on the soil-water variable only (no biomass dependence as in the K model); the dependence is linear at small soil-water contents and approaches a constant value at high contents, representing full plant turgor.
Biomass growth is also affected by mortality ($-\mu b$) and by seed dispersal or clonal growth ($\nabla^2 b$). Soil-water content (equation \eqref{Rw}) is increased by the infiltration of surface water ($Ih$). The biomass dependence of the infiltration rate, $I=I(b)$, captures the infiltration contrast that exists between bare soil (low infiltration rate) and vegetated soil (high infiltration rate) for $f<1$.  The other terms affecting the dynamics of the soil water represent loss of water due to evaporation and drainage ($-\nu w$), water uptake by the plants ($-\gamma Gb$), and moisture diffusion within the soil.
The dynamics of surface water (equation \eqref{Rh}) are affected by precipitation at a rate $p$, by water infiltration into the soil, and by overland flow modeled as a diffusion process.

The R model captures an important pattern-forming feedback - the infiltration feedback. When the infiltration contrast is high ($f\ll 1$)  patches with growing vegetation act as sinks for runoff water. This accelerates the vegetation growth, sharpens the infiltration contrast and increase even further the soil water content in the patch areas. The water flow towards vegetation patches inhibits the growth in the patch surroundings thereby promoting vegetation pattern formation. The infiltration feedback allows vegetation pattern formation at lower, more realistic, values of the soil-water diffusion constant in comparsion to the K model.

\subsection{Gilad et al. (G) model}\label{sec:Meron}
The fourth model to be studied, the G model, was introduced by Gilad et al. \cite{Gilad2004prl,Gilad2007jtb} and contains the same three dynamical variables, $b$, $w$ and $h$ as the R model, but with the interpretation of $b$ as representing the \emph{above-ground} biomass. This is because the G model explicitly considers the root system and the relation between the root-zone size and the above-ground biomass. This additional element allows the introduction of another important pattern-forming feedback besides the infiltration feedback, the root-augmentation feedback. The model equations, in non-dimensional forms, read
\begin{subequations}
 \begin{align}
  \partial_t b&=G_b b\left(1-b\right)-b+D_b\nabla^2 b, \label{Mb} \\
  \partial_t w&=I h-\nu\left(1-\rho b\right)w-G_w w +D_w\nabla^2 w, \label{Mw} \\
  \partial_t h&=p-I h+D_h\nabla^2\left(h^2\right). \label{Mh}
 \end{align}
\end{subequations}
Like in the K model, the biomass growth rate, $G_b$, depends both on $w$ and $b$ but in a non-local way that accounts for the contribution of soil-water availability at point ${\bf x'}$ to biomass growth at point ${\bf x}$ through a biomass-dependent root system that extends from point ${\bf x}$ to point ${\bf x'}$. Similarly, the water-uptake rate, $G_w$, by the plants' roots depends on $b$ and $w$ in a nonlocal manner to account for the uptake at a point ${\bf x}$ by a plant located at ${\bf x'}$ whose roots extend to ${\bf x}$. Specifically,
\begin{subequations}
 \begin{align}
  G_b&=\nu{\displaystyle{\int\limits_\Omega}}g\left({\bf x},{\bf x'},t\right)w\left({\bf x'},t\right)\,, \label{Gb} \\
  G_w&=\gamma{\displaystyle{\int\limits_\Omega}}g\left({\bf x'},{\bf x},t\right)b\left({\bf x'},t\right)\,, \label{Gw} \\
  g\left({\bf x},{\bf x'},t\right)&=\frac{1}{2\pi}\exp\left[-\frac{\left({\bf x}-{\bf x'}\right)^2}{2\left(1+\eta b\left({\bf x},t\right)\right)^2}\right]\,. \label{g}
 \end{align}
\end{subequations}
The root-augmentation feedback is captured by letting the width of the root kernel $g$, which represents the lateral root-zone size, to linearly increase with the above-ground biomass. As a plant grows its root zone extends to new soil regions. As a result the amount of water available to the plant increases and the plant can grow even
further. While accelerating the local plant growth, this process also depletes the soil-water content in the plant surroundings, thereby inhibiting the growth there and promoting vegetation pattern formation. The proportionality parameter $\eta$ appearing in equation \eqref{g} controls the strength of the root-augmentation feedback. It is a measure of the root-to-shoot ratio, a characteristic plant trait. Note that the soil-water dependence of the biomass growth term in equation \eqref{Mb} and of the water uptake term in equation \eqref{Mw} is linear. Nonlinear forms, including that used in the R model, have been studied in Ref. \cite{Kletter2009jtb}. Like in the R model, the infiltration feedback appears through the biomass-dependent form of the infiltration rate $I$:
\begin{equation}
 I=\alpha\frac{b+qf}{b+q}\,. \label{I}
\end{equation}
Other differences with respect to the R model involve the introduction of (i) the logistic growth form $b(1-b)$ in equation \eqref{Mb}, which represents genetic growth limitations at high biomass densities (e.g. stem strength), (ii) the biomass-dependent evaporation rate in the soil-water equation \eqref{Mw} (second term on right side) which accounts for reduced evaporation by canopy shading and introduces a local positive water-biomass feedback, and (iii) nonlinear overland flow term in the surface-water equation \eqref{Mh} motivated by shallow water theory~\cite{Gilad2004prl,santillana2010comp-geo}, rather than a diffusion term as in the R model.

\subsection{Simplified Gilad et al. (SG) model}\label{sec:SimpMeron}
The fifth model is a simplified version of the G model, in which the root kernel $g$ is assumed to vary sharply in comparison to $b$ and $w$, and therefore can be approximated by a Dirac delta function. This approximation is suitable for plant species that grow deep roots with small lateral dimensions. The simplified model, denoted SG, reads
\begin{subequations}
 \begin{align}
\partial_tb&=\nu w b\left(1-b\right)\left(1+\eta b\right)^2-b+D_b\nabla^2b , \label{MMb} \\
\partial_tw&=I h-\nu\left(1-\rho b\right)w-\gamma (1+\eta b)^2 w b +D_w\nabla^2 w , \label{MMw} \\
\partial_th&=p-I h+D_h\nabla^2\left(h^2\right). \label{MMh}
\end{align}
\end{subequations}
This version of the model includes the same pattern-forming infiltration feedback as the original model ($I$ is defined the same way it was defined in the G model), but the root-augmentation feedback is modified; water transport towards growing vegetation patches is no longer a result of uptake by the laterally spread roots, but rather a result of soil-water diffusion.

\section{Results of numerical model studies}
\label{sec:results}
The ecological context we consider is water-limited ecosystems in flat terrains exhibiting bistability of a periodic vegetation pattern and bare soil. We will mostly be concerned with initial states consisting of periodic patterns that are locally disturbed to form bare-soil domains. The numerical studies described below are based on numerical continuation methods, used to identify spatially periodic solutions, and on PDE solvers, used to identify stable branches of localized patterns and to follow the dynamics of bare-soil domains. As we will shortly argue, these dynamics crucially depend on the additional stable pattern states, periodic or localized, that the system supports.

There are several properties that all models appear to share: (i) the coexistence of a family of stable periodic solutions, describing vegetation patterns of different wavelengths, with a stable uniform solution that describes the bare-soil state; (ii) bare-soil domains do not expand into patterned domains; (iii) the existence of a stable localized solution describing a single vegetation spot in an otherwise bare soil state. An additional property that is most significant for regime shifts is not shared by all models - multiplicity of stable hybrid states. We use this property to divide the models into two groups, models that do not show multiplicity of stable hybrid states and models that do show such a multiplicity of states.
The two groups display different forms of regime shifts as described below.

\subsection{Models lacking multiplicity of hybrid states}\label{sec:nosnaking}

\begin{figure}
\begin{center}
 \includegraphics[width=0.95\linewidth]{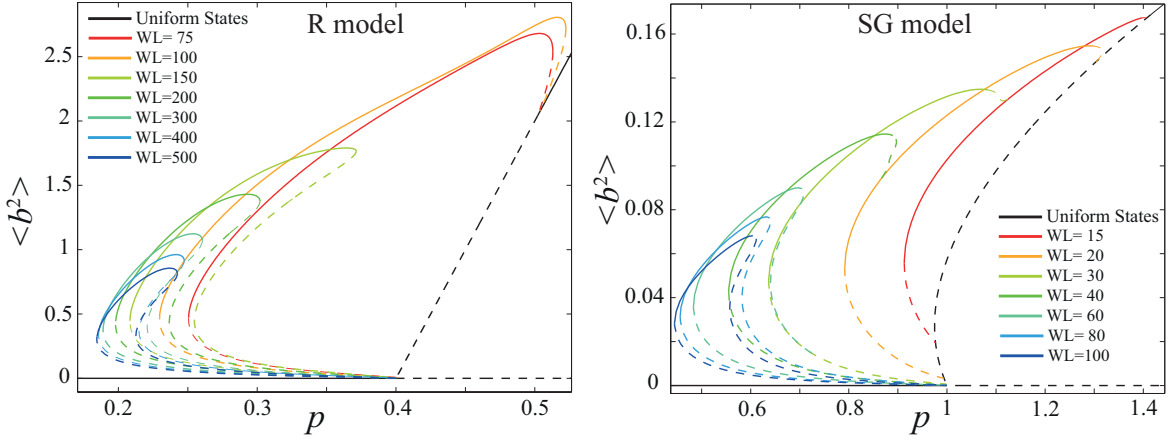}
 \end{center}
 \caption{ Bifurcation diagrams for the R model (left panel) and and the SG model (right panel) in 1D. The diagrams show existence and stability information for uniform vegetation and bare-soil solutions, and for periodic-pattern solutions that differ in their wavelengths (WL) as indicated in the legends; solutions with longer wavelengths extend to lower precipitation values. The vertical axis represents the spatial average of $b^2$ while the horizontal axis represents the precipitation rate. Solid (dashed) lines denote stable (unstable) states. The left most line in both panels corresponds to a single hump (spot). The large overlap ranges of the periodic-pattern solutions allows the system to respond to local disturbances or precipitation changes by changing the pattern's wavelength (see Fig. \ref{fig:localdisturbanceriet}). Hybrid states
 resulting from front pinning were not observed in these models. The diagram for the SG model shows period-doubling bifurcations which were not found in the R model (e.g. the point where the green line WL=30 emanates from the red line WL=15). The instability of a solution that goes through period doubling is not captured (as the solid line indicates) because of the small system considered in the numerical stability analysis. The parameters used for the R model are: $\mu=0.5$, $\alpha=0.4$, $f=0.2$, $\gamma=0.1$, $\nu=0.4$, $D_w=1$, $D_h=1000$. Those for the SG model are: $q=0.05$, $\nu=3.33$, $\alpha=33.33$, $\eta=3.5$, $\gamma=16.66$, $D_w=100$, $D_h=10000$.}
\label{fig:multikriet}
\end{figure}

The models that belong to this group are the K model (Section \ref{sec:Klausmeier}), the R model (Section \ref{sec:Rietkerk}) and the SG model (Section \ref{sec:SimpMeron}). These models have wide bands of periodic solutions with stable branches that coexist with the stable branch of the bare-soil solution~\cite{Dijkstra2011IJBC,vanderStelt2012nonl_sci}. Figure \ref{fig:multikriet} shows bifurcation diagrams for the R and SG models in 1D, computed by a numerical continuation method \cite{auto}. The bifurcation parameter was chosen to be the precipitation rate $p$. The diagrams show overlapping periodic solutions whose wavelengths increase as $p$ decreases. The last periodic solution to exist corresponds to a single hump. We have not been able to identify (by numerical continuation) solution branches that describe hybrid patterns, either groups of humps in an otherwise bare soil state or holes in an otherwise periodic pattern. To further test whether such solutions can exist in these models or, if they exist, whether they are stable, we solved the models' equations numerically using initial conditions that describe fronts separating the patterned and the bare-soil states. Convergence to front solutions that are stationary over a range of $p$ values would indicate the possible existence of hybrid solutions~\cite{Pomeau1986pd, Knobloch2008nonlinearity}. Such front pinning, however, has not been observed; in all simulations the patterned state propagated into the bare-soil state. We conclude that stable hybrid solutions, apart from a single hump solution,
do not exist in these models, or if they do, their existence range is extremely small.

\begin{figure}
 \includegraphics[width=0.83\linewidth]{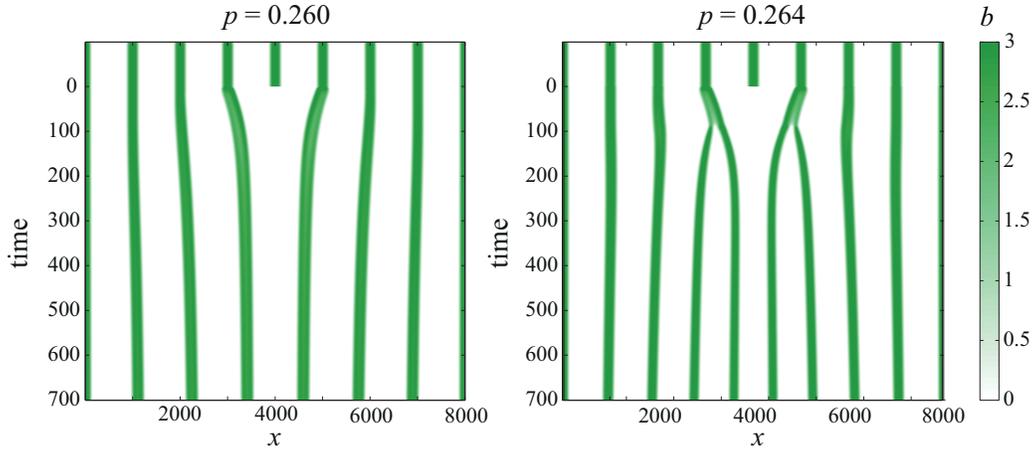}
 \caption{ The response of periodic 1D patterns to local disturbances at different precipitation values in the R model. Shown are space-time plots for $p=0.26$ (left panel) and for $p=0.264$ (right panel). At the lower precipitation value the removal of a hump leads to a pattern with a longer wavelength (the number of humps after the disturbance remains the same and the distance between them is adjusted to fill the whole space). At the higher precipitation the removal of a hump leads to a pattern with a smaller wavelength (the two humps adjacent to the disturbed location split and the number of humps in the final pattern is larger than in the initial pattern); after the splitting the distance between the humps is adjusted to fill the whole space with evenly spaced humps.
 Parameters are as in Fig. \ref{fig:multikriet}.}
\label{fig:localdisturbanceriet}
\end{figure}

\begin{figure}
 \includegraphics[width=0.9\linewidth]{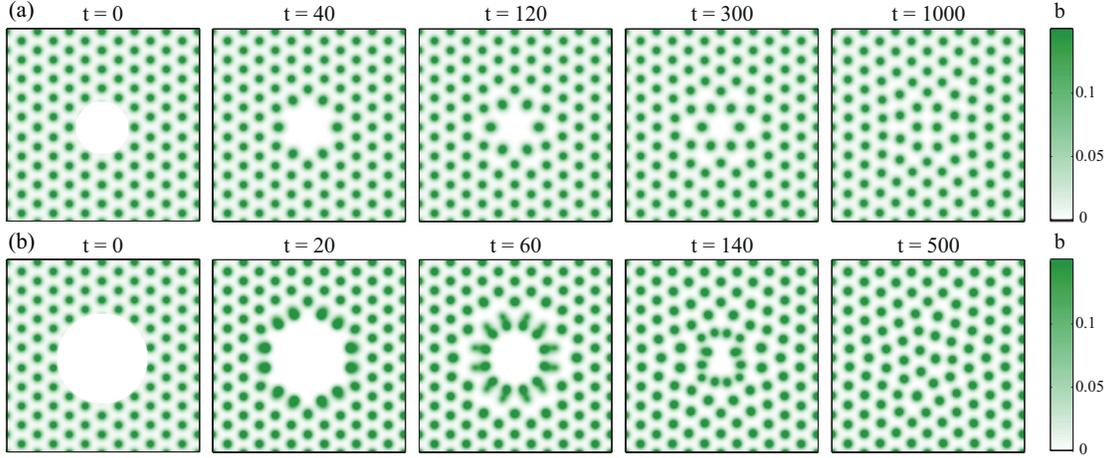}
 \caption{ The response of periodic spot patterns to disturbances of different sizes in the SG model under the same environmental conditions. A small-size disturbance (snapshots in row (a)) leads to a pattern with a longer wavelength through space filling by inter-spot distance adjustments with no change in the total spot number. A large-size disturbance (snapshots in row (b)) is followed by space filling through spot splitting and inter-spot distance adjustments, which generally will result in a change of the total spot number. Both processes can be viewed as a front propagation problem involving a wavenumber change in the pattern left behind the front. The parameters used are: $p=0.9$, $q=0.05$, $\nu=3.33$, $\alpha=33.33$, $\eta=3.5$, $\gamma=16.66$, $D_w=100$ and $D_h=500$.}
\label{fig:localdisturbancegilad}
\end{figure}

\begin{figure}
 \includegraphics[width=0.9\textwidth]{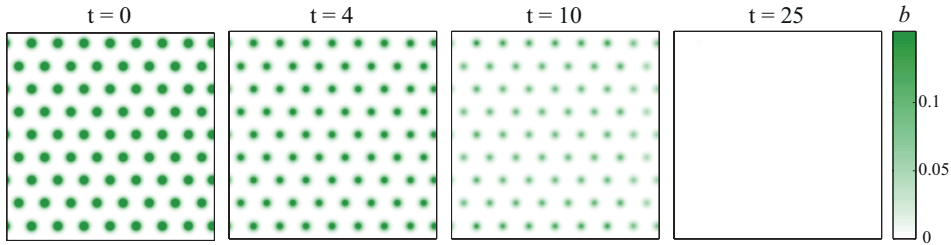}
 \caption{An abrupt transition from a spot pattern to bare soil in the SG model following a precipitation decrease below the bistability range of periodic spot patterns and bare soil. All parameters are as in Fig. \ref{fig:localdisturbancegilad} except for the precipitation which is $p=0.45$.
 }
 \label{fig:simp_gilad_decay}
\end{figure}

In order to study regime shifts in the K, R and SG models we simulated the model equations within the bistability range of periodic patterns and bare soil, starting with periodic patterns that contain bare-soil domains. Since the patterned state was always found to propagate into the bare-soil state, such initial bare-soil domains contract and disappear. This behavior rules out the occurrence of a gradual regime shift to the bare-soil state (similar to that shown in panels e-h of Fig. \ref{fig:abrupt_vs_gradual}). The final pattern, however, can differ from the initial one in its wavelength as the 1D simulations of the R model displayed in Fig. \ref{fig:localdisturbanceriet} show. The system can respond by mere readjustment of the spacings between individual humps without a change in their number, which leads to an increase in the pattern's wavelength (left panel), or, at higher precipitation, by hump splitting, which results in a decrease of the pattern's wavelength (right panel). Similar responses to local
disturbances were found in the K and SG models. Figure \ref{fig:localdisturbancegilad} displays results of 2D simulations of the SG model showing that the two response forms, spacing readjustments and spot splitting, can occur at the same precipitation by changing the size of the initial bare-soil domain. Reducing the precipitation rate to values below the bistability range of periodic patterns and bare soil leads to an abrupt global transition to the bare-soil state as Fig. \ref{fig:simp_gilad_decay} shows.

\subsection{Models exhibiting multiplicity of hybrid states}\label{sec:snaking}
\begin{figure}
 \includegraphics[width=0.9\textwidth]{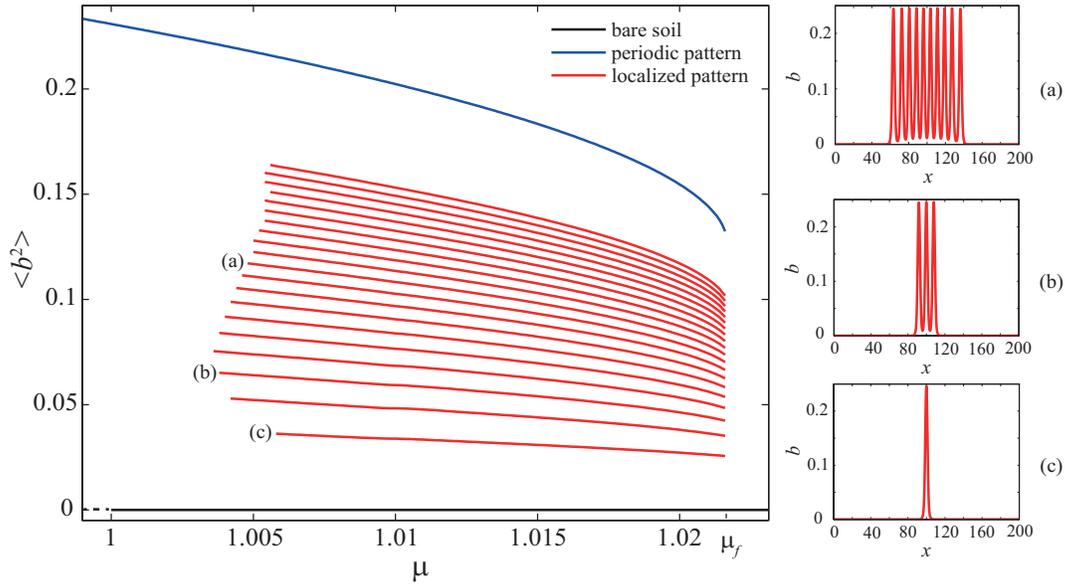}
 \caption{A bifurcation diagram showing hybrid states in the LL model. The vertical axis represents the spatial average of $b^2$ while the horizontal axis represents the mortality rate.
 The top, blue line represents a periodic-pattern state, while the bottom, black line represents the bare-soil state. The red lines in between correspond to localized hybrid patterns with odd and even number of humps as the examples in
 the panels on the right side show.  Note that all hybrid-state branches (red lines) terminate at the same parameter value, $\mu_f$, as the periodic pattern branch. This feature has repeatedly been found for other sets of parameter values and implies an abrupt shift to the bare-soil state upon increasing $\mu$. The parameters we used are: $\Lambda=1.2$ and $L=0.2$.}
 \label{fig:Lejeune_snaking}
\end{figure}

Numerical solutions of the LL and G models (Sections \ref{sec:Lejeune} and \ref{sec:Meron}) using PDE solvers point towards the existence of stable hybrid states in addition to periodic-pattern states~\footnote{The application of numerical continuation methods for these models is harder in comparison to the K, R and SG models and has not been pursued in this study.}.
Fig. \ref{fig:Lejeune_snaking} shows a bifurcation diagram for the LL model, using the mortality rate $\mu$ as the bifurcation parameter. The upper solution branch corresponds to a periodic-pattern state, while the lowest branch corresponds to the bare-soil state. The red branches in between correspond to stable hybrid states describing localized patterns, a few examples of which are shown in the right panels. Solutions of this kind in 1D and 2D have been found earlier~\cite{Lejeune2002pre}. Figure \ref{fig:Gilad_snaking} shows a partial bifurcation diagram for the G model in 2D. The upper line corresponds to a spot-pattern state~\footnote{The spot pattern is rhombic rather than hexagonal. Such patterns apparently exist in the G model, like in other pattern-formation models~\cite{Bachir2001epl}. In the present case, the system converged to the rhombic pattern following a disturbance of an hexagonal pattern. Since it contains a direction of denser spots (vertical) along which the water stress is higher,
transitions to hybrid states can be induced upon decreasing $p$ in the absence of local disturbances.}, while the lower lines correspond to hybrid patterns with decreasing number of spots as the right panels show. Note the difference between the hybrid solution branches in the two models; while in the LL model they all terminate at the same control-parameter value $\mu_f$, which coincides with the fold-bifurcation point of the periodic pattern solution, in the G model the hybrid solution branches are slanted~\cite{Dawes2008sjads} - solutions with smaller numbers of spots terminate at lower $p$ values.

The multitude of stable hybrid patterns, i.e. patterns consisting of groups of spots in an otherwise bare soil, groups of holes in otherwise periodic patterns and various combinations thereof, suggests a form of spatial plasticity. That is, any pattern of local disturbances shifts the system to the closest hybrid pattern with no further dynamics. This behavior rules out the occurrence of a gradual regime shift as a result of initial local disturbances, but unlike the K, R and SG models the system does not recover from the disturbances. This suggests the possible occurrence of a gradual regime shift in a continuously disturbed system.

While the two models share spatial plasticity in response to local disturbances, they differ in the response to gradual parameter changes ($p$ or $\mu$). In the LL model all localized pattern solutions terminate at the fold bifurcation point $\mu_f$ (see Fig. \ref{fig:Lejeune_snaking}). Above that point the only stable state is bare soil and, therefore, any hybrid state must collapse to this state. Note the difference between the bifurcation diagram in Fig. \ref{fig:Lejeune_snaking} and
the diagram obtained with the Swift-Hohenberg equation in Fig. \ref{fig:snaking}. In the latter there is a subrange ($\lambda_f<\lambda<\lambda_1$) outside the hybrid-state range which is still within the bistability range, where disturbed patterns go through gradual shifts. No such subrange has been found in the LL model.
Contrary to the LL model, the slanted structure of localized pattern solutions in the G model, allows for a gradual response. In fact, the hybrid state (b) in Fig. \ref{fig:Gilad_snaking} was obtained from the periodic state (a) by an incremental decrease of $p$. Likewise, the hybrid states (c) and (d) were obtained from the states (b) and (c) by further incremental decreases of $p$. The degree of slanting increases as the root-to-shoot parameter $\eta$ is increased.

\section{Discussion}\label{sec:summary}

\begin{figure}
 \includegraphics[width=0.9\linewidth]{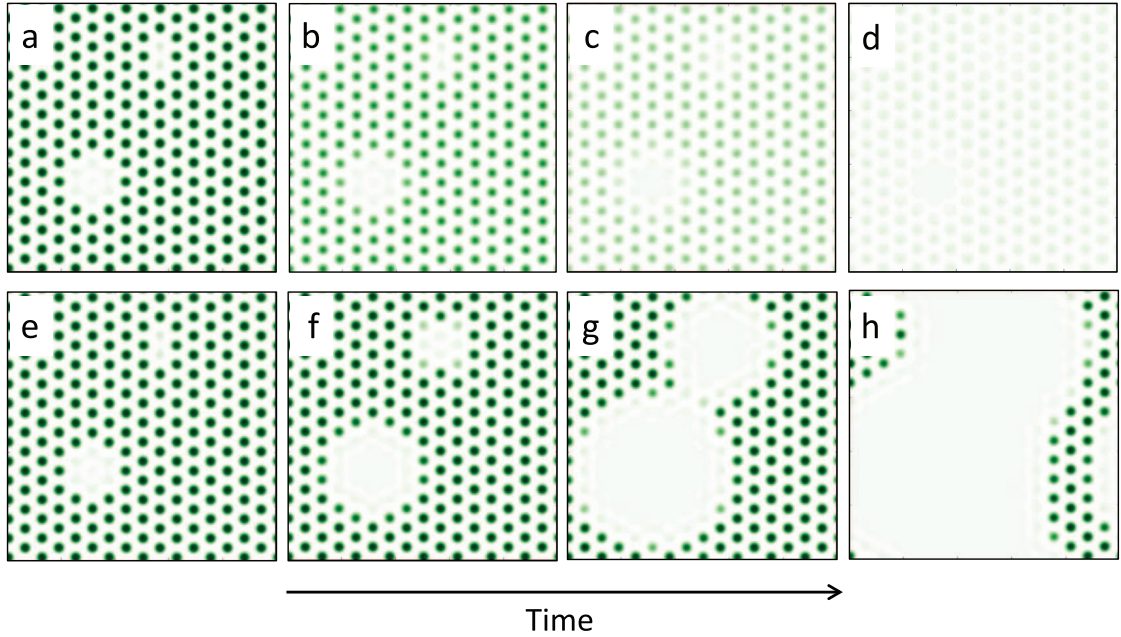}
 \caption{ Illustration of abrupt vs. gradual global regime shifts. Panels (a-d) show an abrupt transition from a
disturbed pattern state (a) to a zero uniform state, occurring globally on a short time scale by decreasing the control parameter below the bistability range, i.e. $\lambda<\lambda_f$ in Fig. \ref{fig:snaking}. Panels (e-h) show a gradual transition from the same initial state to the zero state, within the bistability range, but outside the hybrid-state range, i.e. $\lambda_f<\lambda<\lambda_1$ in Fig. \ref{fig:snaking}. The gradual transition occurs by the local expansion and coalescence of the disturbed domains on a time scale much longer than that of the abrupt transition (note that the latter is so fast that no noticeable domain
expansion occurs during the whole transition). Both shifts are global in the sense that they culminate in a zero state encompassing the whole system (panel (h) is still a transient). The transitions were obtained by solving numerically the Swift-Hohenberg equation~\cite{Knobloch2008nonlinearity,Bel2012theo_ecol}.}
\label{fig:abrupt_vs_gradual}
\end{figure}

All models considered in this study predict the same basic vegetation states and stability properties along a rainfall gradient, including a bistability range of bare soil and periodic spot patterns. We may therefore expect these models to depict similar scenarios for desertification shifts, i.e. transitions from productive spot patterns to the unproductive bare-soil state. Pattern-formation theory,
represented here by results obtained with the Swift-Hohenberg equation\footnote{We refer here to the Swift-Hohenberg equation as a prototype of pattern-formation behaviors in systems showing bistability of uniform and patterned states, and to Ref.~\cite{Bel2012theo_ecol} for the implications of these general behaviors to the context of regime shifts. Similar pattern-formation behaviors have been found with many other pattern-formation models~\cite{Knobloch2008nonlinearity}.}, suggests various possible forms for such scenarios; abrupt, gradual or incipient, induced by environmental changes, by disturbances or both. Underlying these forms are several nonlinear behaviors. The first and simplest is a global transition from the spot-pattern state to the bare-soil state, induced by a slow change of a control parameter past a fold bifurcation, or by a disturbance that shifts the system as a whole to the attraction basin of the bare-soil state. Such processes induce global abrupt shifts to the bare-soil state as Fig. \ref{fig:abrupt_vs_gradual}(a-d) illustrates. Local disturbances, on the other hand, can lead to partial shifts that result in spatially-limited domains of the bare-soil state in an otherwise periodic-pattern state. The subsequent course of events depends on the dynamics of the fronts that bound these domains. When the fronts propagate, a slow process of expansion and coalescence of bare-soil domains can eventually culminate in a global gradual shift, as Fig. \ref{fig:abrupt_vs_gradual}(e-h) illustrates. When the fronts are pinned, the domains remain fixed in size, after some small adjustments, in which case the shift is incomplete or incipient - the system converges to one of the many hybrid states it supports. To our surprise the models we studied do not capture all possible scenarios pattern-formation theory allows. Moreover, scenarios that are captured by some models are not captured by others.

\begin{figure}
\includegraphics[width=0.9\textwidth]{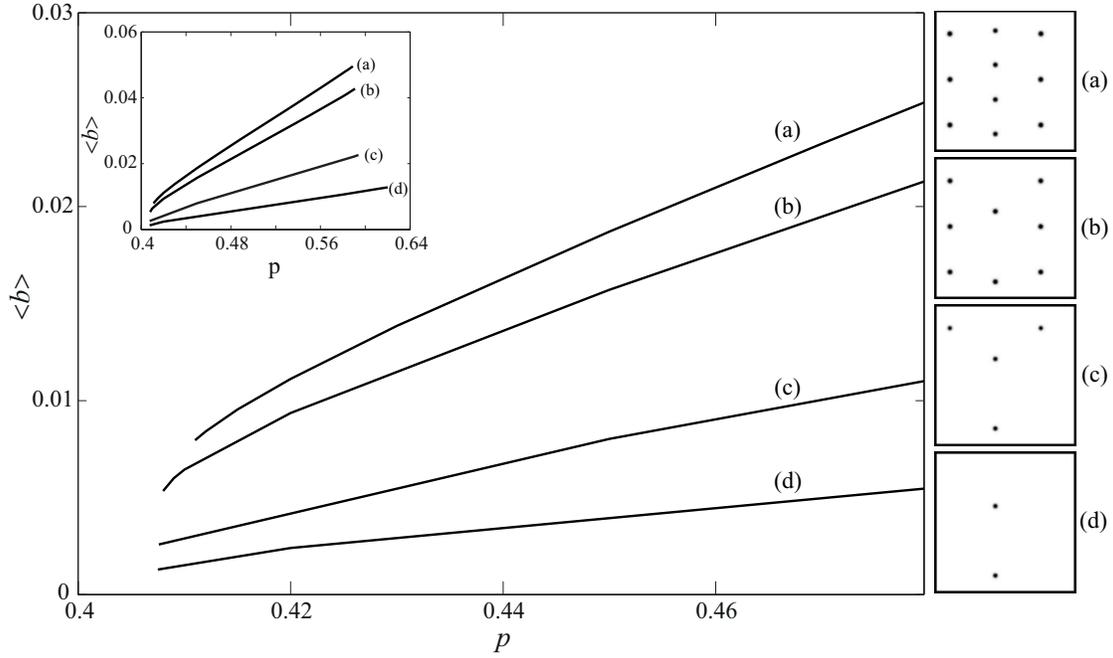}
\caption{A bifurcation diagram showing a few hybrid states in the G model. The vertical axis represents the spatial average of the biomass while the horizontal axis represents the precipitation rate. The upper branch (a) represents a rhombic spot pattern shown in the corresponding panel on the right. Incremental precipitation decrease leads to the local disappearance of spots and the convergence to a stable hybrid state (solution branch and panel (b)). Further incremental decreases lead to hybrid states of lower bioproductivity (solution branches and panels (c,d)). Gradual shifts of this kind towards the bare-soil state are possible because of the slanted structure of the branch edges. The inset shows the full precipitation range in which hybrid states exist or are stable,
and that the high-precipitation edge of the hybrid-state range is also slanted. The backward slanting implies abrupt recovery.
The model parameters used here correspond to strong infiltration contrast ,$f=0.1$ and moderate ``root to shoot'' ratio, $\eta=3$.
The other parameters are: $D_b=0.02$, $D_w=2$, $D_h=200$, $\nu=4$, $\rho=1$, $\gamma=5$, $\alpha=160$ and $q=0.05$. }
 \label{fig:Gilad_snaking}
\end{figure}

Our studies first suggest that in all
five vegetation models (K, R, SG, LL, G) the bare-soil state never grows at the expense of the periodic-pattern state (unlike the behavior shown in Fig. \ref{fig:abrupt_vs_gradual}(e-h)) through the entire bistability range; bare-soil domains either stay fixed in size or contract and disappear. Furthermore,
the K, R, and SG models do not show hybrid states at all,
while the models that do show hybrid states, LL and G, differ in the existence ranges of these states. In the LL model the branches of all hybrid states terminate at the same threshold which coincides with that of  the periodic pattern state, while in the G model the termination points are aligned on a slanted line. The results for the K, R and SG models suggest that shifts to the bare-soil state can only occur outside the bistability range of vegetation patterns and bare soil, and are therefore abrupt. Within the bistability range, bare-soil domains induced by local disturbances contract and disappear, thus restoring the vegetation-pattern state, although a wavelength change is likely to occur. Both the LL and G models predict the possible occurrence of incipient regime shifts within the bistability range of periodic vegetation patterns and bare soil. These shifts can be induced by local-disturbance regimes and culminate in one of the stable hybrid states when the disturbance regimes are over. Complete
shifts to the bare-soil state, due to increased stress, are abrupt in the LL model but can be gradual in the G model because of the slanted structure of the hybrid solution branches; incremental precipitation decrease in the G model can result in step-like transitions to hybrid states of lower bioproductivity as Fig. \ref{fig:Gilad_snaking} shows.

These results raise several open questions. The first is related to the finding that bare-soil domains do not expand into patterned domains in the entire bistability range. This behavior can be attributed to the positive pattern-forming infiltration and root-augmentaiton feedbacks. Both give advantage to plants at the rim of a patterned domain as compared with inner plants; the rim plants receive more runoff from the surrounding bare soil and experience weaker competition for soil water. These factors act against the retreat of vegetated domains. Processes that may favor such a retreat include soil erosion and roots exposure in sandy soils under conditions of high wind power~\cite{Okin2001jae},
or insect outbreak~\cite{Allen2010fem}. Whether bare-soil expansion can be explained by water-biomass interactions alone, or additional processes must be considered, is still an open question that calls for both empirical and further model studies. From the perspective of pattern formation theory the finding that the bare-soil state never expands into vegetation-pattern states questions the utility of the Maxwell-point concept far from the instability of uniform vegetation to periodic patterns and calls for further mathematical analysis.

Another open question is what elements in the LL and G models, and correspondingly what ecological and physical processes, are responsible for the multitude of stable hybrid states. The results for the LL model clearly show that reducing local facilitation, by decreasing the parameter $\Lambda$, narrows down the hybrid-state range and can eliminate the hybrid states altogether. However, it also narrows down the bistability range of periodic patterns and bare soil, and therefore does not resolve processes that favor the formation of localized patterns alone. The results for the G model and its simplified version SG hint towards the possible role of the nonlocal water uptake by laterally extended root systems in inducing hybrid states. This nonlocal competition mechanism is absent in the SG model and may possibly be responsible for the absence of hybrid states in this model.
Further studies are needed, first to substantiate the existence of stable hybrid states, particularly in the G model, and second to clarify the roles of local and nonlocal facilitation and competition processes  in inducing them.

Finally, the models we have studied are all deterministic. Real ecosystems, however, are generally subjected to stochastic fluctuations in time and space, which may affect the bifurcation structure of spatial states. Additive temporal noise, for example, can induce the propagation of pinned fronts~\cite{Clerc2005prl}, and thereby affect the hybrid-state range. The effect of noise on abrupt, gradual and incipient regime shifts is yet another open problem that calls for further studies.

Studying these questions is significant for identifying the nature of desertification shifts, i.e. whether they are abrupt, gradual or incipient, in various environments and for different plant species, and for assessing the applicability of early warning signals for imminent desertification~\cite{Scheffer2009nature}.

\begin{acknowledgments}
We wish to thank Arjen Doelman and Moshe Shachak for helpful discussions and Arik Yochelis for helping with the numerical continuation analysis. The research leading to these results has received funding from the European Union Seventh Framework Programme (FP7/2007-2013) under grant number [293825].
\end{acknowledgments}

\newpage

\end{document}